\documentstyle[11pt,epsfig]{article}
\title{\bf The solitary solutions of
  nonlinear Klein-Gordon field \\with minimal length}

\author{A. Jahangiri$^1$, S. Miraboutalebi$^1$\footnote{E-mail:~s$_{-}$mirabotalebi@iau-tnb.ac.ir}, F. Ahmadi$^2$ and A. A. Masoudi$^{1,3}$ \\
$^1${\small Department of Physics, Islamic Azad University, North Tehran Branch, Tehran, 1651153311, Iran.}\\
$^2${\small Department of Physics, Shahid Rajaee Teacher Training University, Tehran 16788, Iran.}\\
$^3${\small Department of Physics, Alzahra University, Tehran 1993891167, Iran.}\\}
\begin{document}
\maketitle
\begin{abstract}
The existence of a minimal length is predicted by theories of quantum gravity and it is generally accepted that this minimal length should be of the order of the Planck length and hence can be observed in high energy phenomenon. We study the implications of the presence of the minimal length on the Klein-Gordon filed with $\phi^4$ self-interaction. Considering the process of spontaneous symmetry breaking, the potential also includes the $\phi^3$ term. The consequent field equation is a fourth-order differential equation and is considered to have solitary solutions. The $sech$ method is applied and the normalized solutions are obtained in closed forms and the energy spectrum of the solitary fields is determined. The modification parameter of the theory is estimated by the width and the energy of the obtained solitary fields.

\end{abstract}



\section{Introduction}

Incorporating a plausible minimal length into quantum theory demands the correction of the usual Heisenberg uncertainty principle \cite{Hossenfelder}-\cite{Garay}. This subject attracts much attention and seems to be relevant especially at high energies. \cite{Das1}-\cite{Pikovski}. Studies in this area include a wide range of physical sciences from cosmological scales to quantum physics. By considering the minimal length, the observable quantities of a physical system undergo some correction. The induced modifications are assumed to be small enough that these effects are discussed and written in a perturbative manner \cite{Brau}-\cite{Mir4}. However, some creative approaches tend to incorporate more exact results \cite{Chang}-\cite{Hassanabadi}.

In the presence of the minimal length, the uncertainty relation is corrected in the simplest form
\begin{equation}\label{I1}
\triangle x \triangle p\geq \frac{\hbar}{2}\left[1+3\beta(\triangle p)^2+\tau \right]\,,
\end{equation}
where $\tau$ is a function generally independent of $\triangle p$ and $\triangle x$.  This relation indicates the existence of the minimal length  $\Delta X_{min}\approx \hbar\sqrt{3\beta}$. By the assumption $\beta=\beta_0(\frac{\ell_P}{\hbar})^2$, the minimal length becomes $\Delta X_{min}\approx \sqrt{3\beta_0}~\ell_p$ \cite{Mir1}. Here, $\ell_P$ is the Planck length and $\beta_0$ denotes a dimensionless constant.

The modified uncertainty Eq.(\ref{I1}) can be inferred from the deformed algebra introduced in \cite{Kempf1}-\cite{Kempf3}. This algebra has been generalized to a Lorentz covariant form in \cite{Quesne1} and \cite{Quesne2}. The commutation relations of this generalized version is given by
\begin{equation}\label{I2}
[X^{\mu},P^{\nu}]=-i\hbar \left[(1-\beta P^2)g^{\mu\nu}-2\beta P^{\mu}P^{\nu}\right]\,,
\end{equation}
$$
[X^{\mu},X^{\nu}]=0\,,\,\,\,[P^{\mu},P^{\nu}]=0\,,
$$
where, $P^2=P^{\mu}P_{\mu}=(P^{0})^2-|\vec{P}|^2=\frac{E^2}{c^2}-|\vec{P}|^2$.  Eq.(\ref{I2}) is in fact a simplified version of the more general algebra of \cite{Quesne1} and \cite{Quesne2}, where terms of second order in $\beta$ are omitted and the position operators commute \cite{setare}. Using Eq.(\ref{I2}), the uncertainty relation, in $(1+1)$ dimensions, becomes
\begin{equation}\label{I3}
\triangle X \triangle P\geq \frac{\hbar}{2}\left[1+3\beta(\triangle P)^2-\beta\langle(P^0)^2\rangle+\tau \right]\,.
\end{equation}
This uncertainty relation indicates the existence of the following minimal length

\begin{equation}\label{I4}
\Delta X_{min}\approx \hbar \sqrt{3\beta(1-\beta \frac{E^2}{c^2})}\,,
\end{equation}
which reduces to $\Delta X_{min}\approx \hbar\sqrt{3\beta}$, when one only keeps the first order terms in $\beta$.

The commutation relation Eq.(\ref{I2}) can be obtained by the following representation
\begin{equation}\label{I5}
X^{\mu}=x^{\mu}\,,\,\,\, P^{\mu}=(1-\beta p^2)p^{\mu}\,.
\end{equation}
where $x^{\mu}$ and $p^{\mu}$ are the ordinary position and momentum operators satisfying $[x^{\mu},p^{\nu}]=-i\hbar g^{\mu\nu}$.

Eq.(\ref{I3}) represents a relativistic generalized uncertainty principle (RGUP) which can be used to examine the consequences of the minimal length hypothesis in relativistic quantum theory. Among the researches devoted to studying the implications of RGUP of the above-mentioned form,  one can consider \cite{setare} and \cite{Kothawala}. In \cite{setare} the modification of the Klein-Gordon equation in $3+1$  dimensions in the presence of this RGUP is studied. Considering the plane wave solutions, this work shows that the equation naturally describes two massive particles with different masses. This approach also establishes an upper limit on the modification parameter $\beta$ of the theory. In \cite{Kothawala} the RGUP Eq.(\ref{I3}) is applied to a free scalar field and propagation of the field is studied. This work studies two modes $f_{\pm}=1\pm \beta p^2$ in Eq.(\ref{I2}). Applying a three-dimensional spatial Fourier transformation to the classical phase space variables, the RGUP can be written in $k$-space. The propagator of the scalar field for both modes $f_{\pm}$ is obtained in terms of Green functions. This study shows that at short wavelengths only the $f_{+}$ mode leads to the proper propagating waves, whereas at the long-wavelength regime, both $f_{\pm}$ modes are appropriate.

In \cite{Todorinov} another approach of RGUP has been introduced which is different from Eq.(\ref{I3}). In \cite{Bosso} the predictions of RGUP are studied based on the scattering experiment at high energy scales.  This work formulates the Lagrangian quantum electrodynamics of a complex scalar field minimally coupled to RGUP. The massive scalar field in the context of this version of RGUP is a third-order differential equation while the RGUP of Eq.(\ref{I3}) leads to a fourth-order equation. In this rwork, the lagrangian is formulated by using the Ostrogradsky method, and the corresponding Feynman propagators for the scalar and gauge fields are calculated.

In this paper,  the solitary solutions of the nonlinear Klein-Gordon equation in the context of the RGUP of Eq.(\ref{I3}) is studied. To include the spontaneous symmetry breaking, we also consider the $\phi^3$ potential besides the $\phi^4$ nonlinear self-interacting term. The appearance of the $\phi^3$  term is due to the field condensation around the vacuum and spontaneously breaks the symmetry of $\phi\rightarrow -\phi$. This model is so productive in quantum field theory and has already been studied via perturbation techniques \cite{Kaku}. Here some exact solitary solutions of the model are introduced.

In this work, the approach for solving the derived equation is the same as employed in \cite{Malfliet1}-\cite{Malfliet2} and is known as the  $tanh$-method. The extended $tanh$-method \cite{Wazwaz2, Abdou1} and modified extended $tanh$-function method \cite{Soliman, Abdou2} are also some developed versions of this scheme. The exact traveling wave solutions for nonlinear equations with solitary, periodic, and rational forms can be obtained via these techniques. Also,  this approach can be applied to study nonlinear phenomena in many relevant different subjects. In this regard, one may mention, for example, the nonlinear sine-Gordon equation \cite{Wazwaz3}, nonlinear heat conduction and Burgers-Fisher equations \cite{Wazwaz4},  nonlinear Klein-Gordon equation \cite{Sirendaoreji}-\cite{Wazwaz6},  nonlinear fifth-order KdV equation \cite{Wazwaz7} and finally the nonlinear Schr\"{o}dinger equation with variable coefficients \cite{Zayed}.

This paper is organized as follows:  In section two we present the field theory model considered here. Section three is devoted to applying the $sech$ technique to solve the corresponding equations. Also, the general physical conditions of the renormalization and energy considerations are presented. Section four is devoted to introducing solitary solutions for the symmetric theory in two subsections, each of which displaying the different solitary solutions of the theory. In  \emph{4.1} and \emph{4.2}, the sets of solutions are studied without and with the minimal length effects, respectively. In section five the solitary solution in the presence of a minimal length and in the context of spontaneous symmetry breaking is discussed.  The concluding remarks are presented in the last section.

\section{The Model}

Let us begin our study by considering the following Klein-Gordon field with a $\phi^4$ interaction \cite{Kaku}

\begin{equation}\label{M1}
{\cal L}_o=\frac{1}{2}\partial_{\nu}\phi\partial^{\nu}\phi-\frac{1}{2}\mu^2\phi^2-\frac{1}{3}\lambda v \phi^3-\frac{1}{4} \lambda \phi^4\,.
\end{equation}
Here, $\mu$ and $\lambda$ are real parameters. In this lagrangian density the symmetry breaking term $\phi^3$ appears due to shifting the field $\phi$ around the vacuum
\begin{equation}\label{v}
\phi_0=\frac{v}{3}=\pm\sqrt{\frac{\mu^2}{2\lambda}}\,.
\end{equation}
The conjugate to $\phi$ namely $\pi$ is defined as

\begin{equation}\label{M2}
\pi_o = \frac{\partial {\cal L}_o}{\partial \dot{\phi}}= \frac{1}{c}~ \partial^{0}\phi\,.
\end{equation}
Then the Hamiltonian becomes

\begin{equation}\label{M3}
{\cal H}_o = \dot{\phi}~\pi-{\cal L}_o =  \partial_{0}\phi ~ \partial^{0}\phi- {\cal L}_o\,.
\end{equation}
Applying the Euler-Lagrange equation leads to the following field equation

\begin{equation}\label{M4}
\Box \phi+\mu^2 \phi +\lambda v \phi^2+\lambda \phi^3=0\,,
\end{equation}
where $\Box=\partial_{\nu}\partial^{\nu}$. This equation has already been solved by using perturbation techniques.
Under the assumption of the RGUP in Eq.(\ref{I3}), the momentum operators are changed according to Eq.(\ref{I5}). Therefore, in the position representation, one obtains

\begin{equation}\label{M5}
\partial^{\mu}\longrightarrow \left(1+\beta\hbar^2\Box\right)\partial^{\mu}\,.
\end{equation}
where it has been substitute $p^2=p_{\nu}p^{\nu}=-\hbar^2\Box$. Applying Eq.(\ref{M5}) in Eq.(\ref{M1}), the Lagrangian in the presence of the minimal length effects becomes

 \begin{equation}\label{M6}
{\cal L}={\cal L}_o+{\cal L}_{GUP}\,,\,\,\,
\end{equation}
where ${\cal L}_{GUP}$ is the correction term of the Lagrangian due to the minimal length effects
\begin{equation}\label{M7}
{\cal L}_{GUP}=-\beta~\hbar^2~  \Box\phi \Box\phi\,.
\end{equation}
Also the conjugate $\pi$  becomes
\begin{equation}\label{M8}
\pi=\frac{1}{c}~ \partial^{o}\phi+ \beta \frac{\hbar^2}{2}\Box \partial^{0}\phi\,,
\end{equation}
where the first term is the ordinary conjugate given in Eq.(\ref{M2}) and the second term denotes its correction. Inserting Eq.(\ref{M5}) into Eq.(\ref{M3}), the Hamiltonian becomes

\begin{equation}\label{M9}
{\cal H}={\cal H}_o+{\cal H}_{GUP}\,,
\end{equation}
where ${\cal H}_{o}$ is given in Eq.(\ref{M3}) and the correction term  ${\cal H}_{GUP}$ is defined as

\begin{equation}\label{M10}
{\cal H}_{GUP}=\beta~\hbar^2\left(2\partial_{o}\phi \Box \partial^{o}\phi+\Box\phi\Box\phi~\right)\,.
\end{equation}
Substituting Eq.(\ref{M5}) in field equation Eq.(\ref{M4}) one find

\begin{equation}\label{M11}
\Box \phi+2\beta\hbar^2\Box\Box\phi+\mu^2 \phi +\lambda v \phi^2+\lambda \phi^3=0\,.
\end{equation}
This is the equation that governs the spontaneous symmetry breaking field $\phi$ in the presence of the minimal length.

\section{Applying the $sech-tanh$ approach}
Here, we assume the existence of the solitary solution for the $1+1$ version of Eq.(\ref{M11}), where $\Box=\frac{\partial^2}{c^2\partial t^2}-\frac{\partial^2}{\partial x^2}$. The solitary solutions move undistorted with constant velocity and are traveling waves of the form
\begin{equation}\label{S1}
\phi(x,t)=\psi(\xi)\,,\,\,\,\xi=g(x-wt)\,,
\end{equation}
where, $\psi(\xi)$ represents a single localized and nonsingular wave solution which travels with speed $w$, having a width proportional to $g^{-1}$. Equivalently, any static localized solution of Eq.(\ref{M11}) is also  a solitary wave which by a simple boost can be transformed to a moving coordinate frame \cite{Rajaraman}. One then obtains either an advanced mode $(x-w t)$ or a retarded mode $(x+w t)$, individually. In fact, due to the nonlinearity, a simple linear combination of both of these modes does not represent a proper solution.

To generalize the method applied here to $3+1$  dimensions, one should firstly decouple the radial and angular parts of Eq.(\ref{M11}). Since the radial variable $r$ explicitly appears in the obtained radial equation, one should first study the static case and find the localized proper solutions. Equivalent to Eq.(\ref{S1}), in this case, one takes $\xi=g~r$ and implements the procedure in what follows.

To find the assuming solitary solutions for Eq.(\ref{M11}), we apply the $sech-tanh$ method \cite{Wazwaz1}. Under the transformation given in Eq.(\ref{S1}), Eq.(\ref{M11})  becomes

\begin{equation}\label{S2}
\left(\eta ~\frac{d^2}{d\xi^2}+2~\beta~\hbar^2\eta^2~\frac{d^4}{d\xi^4}+\mu^2+\lambda~ v~ \psi+\lambda~\psi^2\right)\psi=0\,,
\end{equation}
where $\eta$  is a dimensionless constant defined as
\begin{equation}\label{S3}
\eta=g^2\left(\frac{w^2}{c^2}-1\right)\,.
\end{equation}
Let us now define the following dimensionless quantities
\begin{equation}\label{S4}
\left\{ \begin{array}{ll}
  \xi=\ell_P~\tilde{\xi}\,,\,\,\,\, \psi^2=\ell_P~{\cal E}_P~\tilde{\psi}^2\,,\,\,\,\,v=\sqrt{\ell_P~ {\cal E}_P}~\tilde{v}\,,\\\\
  \lambda=\frac{\tilde{\lambda}}{\ell_p^3~ {\cal E}_P}\,,\,\,\,\,\mu^2=\frac{\tilde{\mu}^2}{\ell_P^2}\,,\,\,\,\,\beta=\frac{\beta_0~\ell_p^2}{2~\hbar^2}=\frac{\beta_0}{2~{\cal P}_P^2}\,,
  \end{array}
\right.
\end{equation}
where, $\ell_p$, ${\cal P}_P$ and ${\cal E}_P$ are the Planck length, the Planck momentum, and the Planck energy, respectively. Using Eq.(\ref{S4}),  Eq.(\ref{S2}) can be rewritten as

\begin{equation}\label{S5}
\left(\eta \frac{d^2}{d\tilde{\xi}^2}+\beta_0\eta^2\frac{d^4}{d\tilde{\xi}^4}+\tilde{\mu}^2+\tilde{\lambda}\tilde{ v }\tilde{\psi}+\tilde{\lambda}\tilde{\psi}^2\right)\tilde{\psi}=0\,.
\end{equation}
Let us apply the transformation

\begin{equation}\label{S6}
z=sech(\tilde{\xi})\,,\,\,\,\tilde{\psi}(\tilde{\xi})=\Phi(z)\,,
\end{equation}
where all the derivatives in Eq.(\ref{S5}) get changed accordingly and this equation becomes

\begin{equation}\label{S7}
 \beta_0{\eta}^{2}\left( 1-z^2 \right) ^{2}{z}^{4} {\frac {{\rm d}^{4}}{{\rm d}{z}^{4}}}\Phi
 +12\beta_0\,{\eta}^{2} \left( {z}^{2}-\frac{1}{2}\right)  \left( z^2-1
 \right)  {z}^{3}{\frac {{\rm d}^{3}}{{\rm d}{z}^{
3}}}\Phi +36\,\eta\, \left[  \beta_0 \left( {z}^{4}-{\frac {
10}{9}{z}^{2}}+{\frac{7}{36}} \right) \eta\right.
\end{equation}
$$
\left. -\frac{1}{36}\,({z}^{2}-1)\right] {z}^{2}{\frac {{\rm d}^{2}}{{\rm d}{z}^{2}}}\Phi+24\,\eta\, \left[  \beta_0 \left( {z}^{4}-\frac{5}{6}\,{z}^{2}+\frac{1}{24}
 \right) \eta-\frac{1}{12}\,{z}^{2}+\frac{1}{24}\right] z{\frac {\rm d}{{\rm d}z}}
\Phi+\left( v\lambda\,\Phi
 +\lambda\, \Phi^{2
}+\mu^2 \right)\,\Phi  =0\,.
$$
Now, suppose that the solitary solution of the last equation can be displayed as polynomials in $sech$ functions. Upon substituting the following ansatz into Eq.(\ref{S7}) we find

\begin{equation}\label{S8}
\Phi(z)=a_1\,z+ a_2\,z^2\,,
\end{equation}
where, $a_1$ and $a_2$ are some unknown constants. In this way, Eq.(\ref{S7}) become an 8th-degree equation in $z$. The coefficients of the same powers of $z$ must be zero which establishes eight relations between parameters of the theory. By solving the relations, one can find the appropriate physical parameters. One can also examine the constant term $a_0$ in Eq.(\ref{S8}) as has already been done in the original works, \cite{Wazwaz1}. However, it should be noticed that the solutions with $a_0\neq 0$ do not respect to the normalization condition. Hence, in what follows, we only introduce the solutions in the form of Eq.(\ref{S8}) with $a_0= 0$ which allows normalization.

The normalization condition for the dimensionless field $\tilde{\psi}$ is given as
\begin{equation}\label{S10}
1=\int_{-\infty}^{\infty} d\tilde{\xi}~|\tilde{\psi}(\tilde{\xi})|^2\,.
\end{equation}
The rest energy of the solitary wave is
\begin{equation}\label{S11}
E=\int_{\infty}^{\infty} dx~{\cal H}(x,t=0) \,,
\end{equation}
where ${\cal H}(x,t=0)$ is given by Eq.(\ref{M9}). Considering the transformations Eqs.(\ref{S1}) and (\ref{S4}), the energy can be calculated by using the following relation

\begin{small}
\begin{equation}\label{S12}
E={\cal E}_P \sqrt{\frac{\alpha}{-\eta}}\int_{-\infty}^{\infty} d\tilde{\xi}\left\{\frac{\eta(\alpha-2)}{2~\alpha}
\left(\frac{d\tilde{\psi}}{d\tilde{\xi}}\right)^2+\frac{1}{2}\tilde{\mu}^2\tilde{\psi}^2+
\frac{1}{3}\tilde{\lambda}\tilde{v}\tilde{\psi}^3+\frac{1}{4}\tilde{\lambda}\tilde{\psi}^4+
\beta_0\eta^2\left[\frac{\alpha-1}{\alpha}\frac{d\tilde{\psi}}{d\tilde{\xi}}\frac{d^3\tilde{\psi}}{d\tilde{\xi}^3 }+\frac{1}{2}\left(\frac{d^2\tilde{\psi}}{d\tilde{\xi}^2}\right)^2\right]\right\}\,,
\end{equation}
\end{small}
where $\alpha=1-\frac{w^2}{c^2}$. To find the energy of the solitary field, it only needs to substitute the field $\tilde{\psi}$  and the appropriate parameters of the model in  Eq.(\ref{S12}).

Inserting Eq.(\ref{S8}) into Eq.(\ref{S7}), one find an $8$th-order  equation in $z$. This equation always is satisfied if the collective coefficient of each power of $z$ be zero. In this way, one obtains a set of $8$ relations between the parameters whose solutions is given in the next section.  These solutions are divided into two different categories.  One set contains the solutions without considering the spontaneous symmetry breaking effects $(v= 0)$  and the other set corresponds to the solutions with such effects $(v\neq 0)$. In each state, the solutions are also derived in the absence $(\beta_0=0)$ and presence $(\beta_0\neq 0)$ of the minimal length hypothesis.


\section{Solutions of the symmetric theory}

This situation corresponds to the ordinary massive Klein-Gordon theory with $\phi^4$ nonlinear term and specified by the condition $v=0$.  Consequently, the spontaneous symmetry breaking term vanishes, and the theory is symmetric. After solving the system of $8$-equations obtained according to the procedure explained in the previous section, we find some solitary solutions. In what follows, we observe that the procedure applied here leads to the exact solitary solutions in both cases, in the presence of the minimal length effects, namely for $\beta_0\neq0$, and without considering the minimal length, namely for $\beta_0=0$.

\subsection{The solitary field in the absence of the minimal length, for $\beta_0=0$}

Let us first consider the simplest case characterized by $v=0$ and $\beta_0=0$. In this situation, the theory is symmetric under $\phi\rightarrow -\phi$ and the hypothesis of the existence of the minimal length is not applied. In this case,  the following specific set of solutions is determined

\begin{equation}\label{F1}
\tilde{\mu}^2 = -\eta\,,\,\,\, \tilde{\lambda} = \frac{2\eta}{a_1^2}\,,\,\,\, a_2 = 0\,,\,
\end{equation}
where $a_1$ and $\eta$ are free parameters. Using these parameters, the solitary wave solution of Eqs.(\ref{S8}) and (\ref{S6}) becomes

\begin{equation}\label{F2}
\tilde{\psi}(\tilde{\xi})=a_1~ sech (\tilde{\xi})\,.
\end{equation}
Under the normalization condition Eq.(\ref{S10}), one finds $a_1 =\pm\frac{\sqrt{2}}{2}$, therefore
\begin{equation}\label{F3}
\tilde{\psi}(\tilde{\xi})=\pm ~\frac{\sqrt{2}}{2}~ sech(\tilde{\xi}) \,.
\end{equation}
Also, the parameter $\tilde{\lambda}$ becomes

\begin{equation}\label{la}
\tilde{\lambda}=4~\eta\,.
\end{equation}
Inserting the field Eq.(\ref{F3}) into Eq.(\ref{S12}), the energy is obtained

\begin{equation}\label{F4}
E=g~\left(\frac{10-7~\alpha}{30}\right){\cal E}_P\,.
\end{equation}
Since $\eta$ is a free parameter in this case, from Eq.(\ref{S3}), $g$is a free parameter too.  For $w<c$, the parameter $\alpha=1-\frac{w^2}{c^2}$ becomes $0<\alpha<1$. For positive $g$, from Eq.(\ref{F4}), the solitary field receives a real positive energy. The width of the solitary wave is characterized by $g^{-1}$. Therefore, for a constant value of $\alpha$,  growing $g$ causes  the width of the solitary field to decrease which,  according to Eq.(\ref{F4}),  leads to the solitary field carrying more energy.

\subsection{The solitary field in the presence of the minimal length, for $\beta_0\neq 0$}

The other relevant solution, corresponding to $v=0$, is as follows

\begin{equation}\label{F5}
\tilde{\mu}^2 = -\frac{16~\eta}{5}~\,,\,\,\, a_1 = 0\,,\,\,\, \beta_0 = -\frac{1}{20~\eta}\,, \,\,\,\tilde{\lambda} = \frac{6~\eta}{a_2^2}\,,
\end{equation}
where $a_2$ and $\eta$ are two arbitrary parameters. By these parameters, the solitary wave solution of Eqs.(\ref{S8}) and (\ref{S6}) becomes

\begin{equation}\label{F6}
\tilde{\psi}(\tilde{\xi})=a_2~ \left(sech(\tilde{\xi})\right)^2\,.
\end{equation}
Applying  condition Eq.(\ref{S10}) and using $a_2$  leads to $a_2=\pm\frac{\sqrt{3}}{2}$, with the normalized field  given as

\begin{equation}\label{F7}
\tilde{\psi}(\tilde{\xi})=~\mp\frac{\sqrt{3}}{2} \left(sech(\tilde{\xi})\right)^2\,,
\end{equation}
and the parameter $\tilde{\lambda}$ of the model in Eq.(\ref{F5}) becomes $\tilde{\lambda}= 8~\eta$.
Substituting the parameters of this subsection into Eq.(\ref{S12}), we find

\begin{equation}\label{F8}
E=2~g\left(\frac{16-19 \alpha}{35}\right)~{\cal E}_P\,.
\end{equation}
For $w<c$ one gets $0<\alpha< 1$, hence, the energy $E$ is real and positive in this case. Again, by growing  $g$ the width of the soliton field decreases and the energy $E$ increases for constant $w$.

The modification parameter $\beta_0$, from Eq.(\ref{F5}),  is determined as $\beta_0=(20 g^2 \alpha)^{-1}$, where $\eta$ is replaced by $\eta=- g^2 \alpha$. To study the behavior of $\beta_0$ one needs to  make the unite of the parameters more plausible. We note that the theory considered  here is suitable for describing a meson field. The meson mass is about $100\, MeV $ to $ 10\,GeV $. But the Planck energy is ${\cal E}_P= 1.22 \times 10^{19}\, GeV $. Therefore, let us redefine the parameter $g$ and $\beta_0$ as

 \begin{equation}\label{R4}
g^{'}=g\times1.22\times10^{22}\,,\,\,\,\beta^{'}_0=\frac{\beta_0}{(1.22)^2}\times 10^{-44}\,.
 \end{equation}
By these redefinitions in Eq.(\ref{F8}) the energy $E$ can be written in $MeV$ unites and parameter $\beta_0$ is rewritten as $\beta^{'}_0=(20 g^{'2} \alpha)^{-1}$. Parameter $\beta^{'}_0$  is illustrated in Fig.(1a) as a function of $g^{'}$ and $\alpha$. One expects that at lower velocities $w$, parameter $\alpha$ increases and $\beta^{'}_0$ decreases and vice versa. Hence, by narrowing the solitary waves, $\beta^{'}_0$ decreases.

To compare energies Eqs.(\ref{F8}) with (\ref{F4}), the relative change in energy, in the presence and absence of the minimal length effects, with respect to the ordinary case, is obtained as follows
\begin{equation}\label{F9}
\frac{\Delta E}{E}={\frac {122-179\,\alpha}{70-49\,\alpha}}\,.
\end{equation}
This relative change depends directly on $\alpha$ and is illustrated in Fig.(1b). The figure shows that at higher velocities the relative change in energy is positive and at lower velocities, this rate acquires negative values.

\begin{figure}
\centerline{\begin{tabular}{c}
\epsfig{file=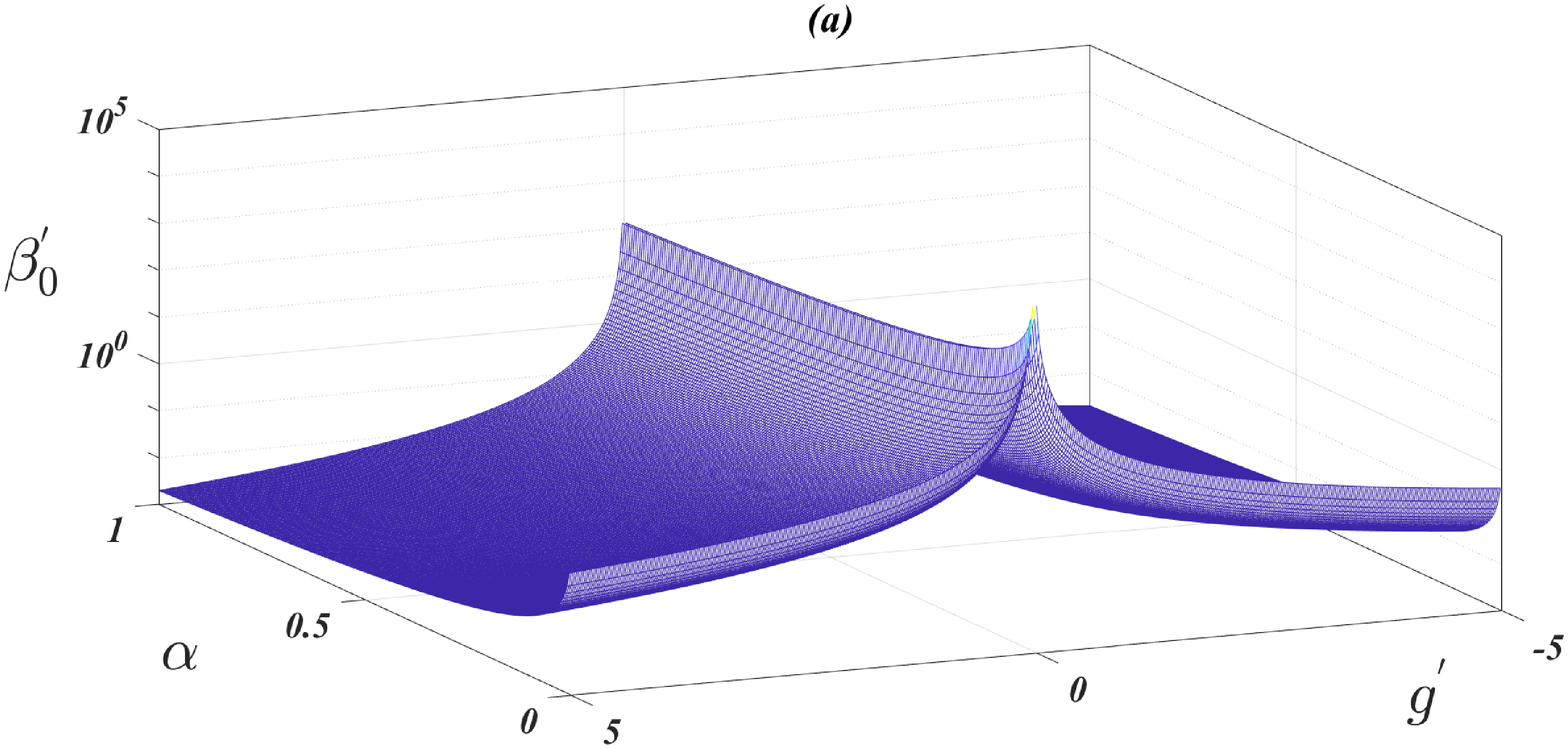,width=8.5cm}
\epsfig{file=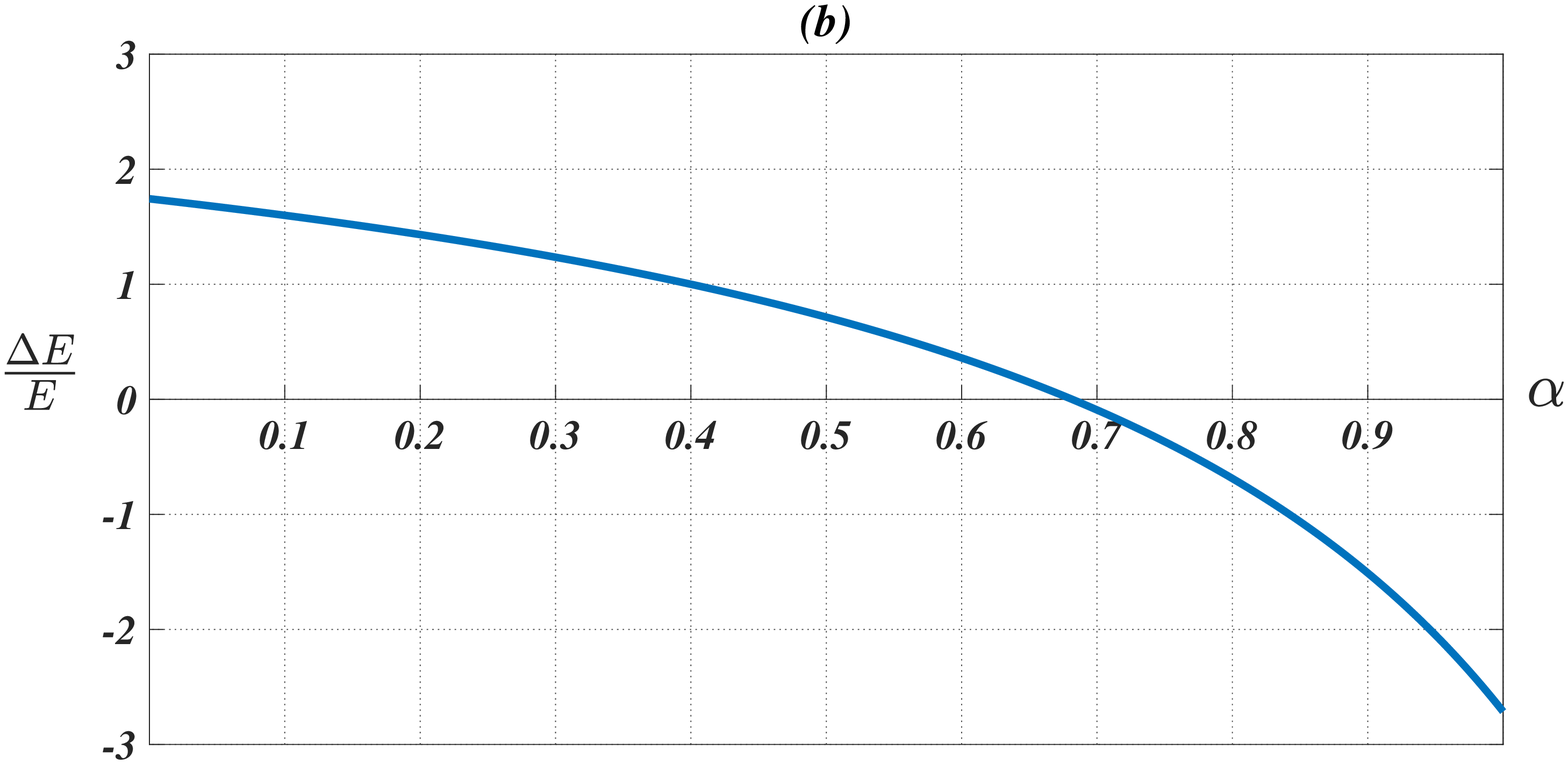,width=8.5cm}
\end{tabular}}
\caption{\small{The parameter $\beta^{'}_0$ as a function of $g^{'}$ and $\alpha$, (a) and the relative change in the energy spectrum $\frac{\Delta E}{E}$ as a function of $\alpha$, (b).}}
\end{figure}

\section{The solutions of the spontaneous symmetry breaking theory}

This situation corresponds to $v\neq 0$, and the symmetry is spontaneously broken. By the method applied here, the minimal length hypothesis plays an important role in determining the exact solitary solutions. One cannot find any solitary solution without the minimal length hypothesis, namely for $\beta_0=0$. However, the solitary solutions can be found in the presence of the minimal length, namely for $\beta_0\neq 0$. Let us first attempt to introduce the following set of solutions
\begin{equation}\label{R1}
\tilde{\mu}^2 = -16\,{\eta}^{2}\beta_0-4\,\eta\,,\,\,\,\tilde{\lambda}=-120\,{\frac {{\eta}^{2}\beta_0}{{{a_2}}^{2}}}\,,\,\,\, \tilde{v}=-{\frac { \left( 20\,\eta\,\beta_0+1 \right) {a_2}}{20\eta\,\beta_0}}
\,,\,\,\, a_1 = 0\,,
\end{equation}
where $a_2$, $\beta_0$  and $\eta$  are the arbitrary parameters. The solitary field in this case, from Eq.(\ref{S8}), using $a_1=0$  and the existence of an arbitrary $a_2$, according to Eq.(\ref{R1}), is the same as the solution of Eq.(\ref{F7}).
Substituting $a_2=\frac{\sqrt{3}}{2}$ in Eq.(\ref{R1}), the parameters $\tilde{\lambda}$ and $\tilde{v}$ become

\begin{equation}\label{R2}
\tilde{\lambda} =-160\,{\eta}^{2}\beta_0\,,\,\,\,\tilde{v}=-\,{\frac { \left( 20\,\beta_0\eta+1 \right) \sqrt {3}}{40\eta\,\beta_0}}\,.
\end{equation}

For the energy, by inserting the corresponding parameters of this subsection into Eq.(\ref{S12}), one obtains

\begin{equation}\label{R3}
E={\frac {8\,g}{7} \left[ \alpha\,\left( \alpha+2 \right) {g
}^{2}\beta_0-\alpha+{\frac{7}{10}} \right] }~{\cal E}_P\,.
\end{equation}
Fig.(2) displays the energy $E$ versus $g^{'}$ and $\beta^{'}_0$ where these two parameters are given by Eq.(\ref{R4}) and also $w=o.5\,c$. Only the positive energies is considered in the figure, hence it includes two different branches. In fact, these two branches correspond to each of the following choices

\begin{equation}\label{w}
\left\{
  \begin{array}{ll}
    g^{'}>0, & \hbox{$\beta^{'}_0>\frac{\left(\alpha-0.7\right)}{g^{'2}\alpha(\alpha+2)}$;} \\
    g^{'}<0, & \hbox{$\beta^{'}_0<\frac{\left(\alpha-0.7\right)}{g^{'2}\alpha(\alpha+2)}$.}
  \end{array}
\right.
\end{equation}
According to the figure, by proper choice of the parameters $g^{'}$ and $\beta^{'}_0$, the solitary wave obtains positive values. However, one can find negative energies as well which shows the presence of the bound state solutions.

\begin{figure}
\centerline{\epsfig{file=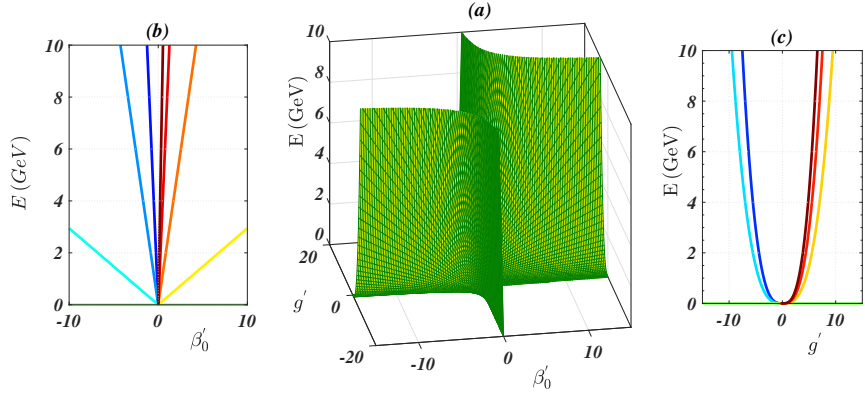,width=12cm}\\
}
\caption{\small{ The energy spectrum $E$ as a function of $g^{'}$ and $\beta^{'}_0$ for $w=0.5~c$.}}
\end{figure}


\section{Conclusions}
In this work we have studied the consequences of the existence of a minimal length on the solutions of a theory with spontaneous symmetry breaking. The hypothesis of the existence of the minimal length comes from quantum gravity and quantum field theory. The theory with spontaneous symmetry breaking considered here is the massive Klein-Gordon with the nonlinear $\phi^4$ and $\phi^3$ terms. Without the $\phi^3$ term, this theory is symmetric and it is well known to have solitary solutions. Here, we assumed that the full theory of spontaneous symmetry breaking in the context of RGUP has solitary solutions. To investigate this assumption,   the most reliable method of obtaining the possible solitary fields, that is the so-called $Sech$ method, was applied and was conducted in two parts, namely the symmetrictheory and the spontaneously broken symmetric theory.  Some solitary solutions were obtained in both situations which show interesting properties.

In the symmetric theory, two different sets of solitary solutions were obtained which correspond, respectively, to solutions with and without considering the minimal length hypothesis. In the absence of a minimal length, the solitary wave is given to first order in $sech$ function. However, the minimal length results in a solitary field which leads to a function being second order in $sech$. In both cases, the rest energy is real and increases by decreasing the width of the solitary fields. For the solutions in the presence of the minimal length, the modification parameter $\beta_0$ was estimated. It was shown that this parameter depends on the energy regime considered. In fact, by increasing the rest energy of the solitary field, the modification parameter decreases and vice-versa. This conclusion is in agreement with other investigations that establish similar relation between the modification parameter and the energy of a physical system.

In the spontaneous symmetry breaking theory the solitary solutions was found only in the attendance of the minimal length hypothesis. By the method applied here, in the absence of the minimal length, the exact solitary solutions were not possible which means that the assumption of the existence of a minimal length is necessary for the solitary solutions. However, some perturbation methods may lead to this kind of solution as well. Here, for the theory with spontaneous symmetry breaking, a real set of solitary solutions to second-order in the $sech$ was found.

\end{document}